# quaint: An R Package for detecting introgression across a phylogeny using discordant gene tree topologies.

Ethan A. Baldwin, James H. Leebens-Mack

Department of Plant Biology, University of Georgia, Athens, Georgia USA

Corresponding author: Ethan Baldwin, ethan.baldwin@uga.edu

**ABSTRACT**

**Premise**

Hybrid speciation and introgressive hybridization are increasingly recognized as important evolutionary phenomena across the tree of life. One widely used class of methods to detect introgression includes D statistics and related methods which employ the ABBA-BABA test using nucleotide site patterns. Recent studies have applied this theoretical framework to phylogenomic datasets using gene tree topologies instead, but no software packages using this method have been developed.

**Methods and Results**

An R package was developed to facilitate the inference of introgression given a set of gene trees and a species tree. Using an ABBA-BABA framework, this package summarizes patterns of gene tree discordance to infer introgression across large phylogenies.

**Conclusions**

Using gene tree topologies, quaint overcomes the limitations of site-based methods, enabling the detection of introgression across broad phylogenomic contexts. This R package provides an accessible and reproducible tool for researchers investigating reticulate evolution.

## INTRODUCTION

Hybridization and introgression between lineages are important evolutionary mechanism for species diversification across the tree of life (Soltis and Soltis 2009;Mallet et al. 2016;Taylor and Larson 2019). Understanding the extent of hybridization can provide insights into the nature of speciation and species boundaries (Harrison and Larson 2014;Gompert et al. 2017). Introgression of genes across species can spawn novel genetic combinations with the potential to drive adaptation (Pease et al. 2016). With the evolutionary significance of hybridization, it is necessary to develop methods for accurately detecting introgression across a wide range of biological contexts.

Currently, some of the most widely used methods for detecting introgression, such as Patterson's D-statistic (Green et al. 2010), examine the frequency of biallelic nucleotide site patterns to detect introgression. These are accurate when examining introgression between closely related species or gene flow between populations from the same species. However, methods based on an infinite-sites model for analyses of biallelic nucleotide site patterns are not well suited for "phylogenomic" datasets, which can include distantly related taxa and a high degree of linkage disequilibrium among sites within a locus. Tight linkage, homoplasy and back mutations, which are more likely to occur given more divergence between organisms, break the assumptions of these site-based methods. Additionally, more divergence leads to technical limitations in producing the data necessary for using site-based methods, such as inabilities to call biallelic single nucleotide polymorphisms (SNPs) against a reference genome and the use of alternative sequencing strategies (i.e. target capture) to ensure the comparison of homologous genes in phylogenomic studies.

A promising alternative to using site-based methods to detect introgression in phylogenomic datasets is to instead compare the frequencies of directly estimated gene tree topologies. Using gene trees instead of nucleotide sites to perform ABBA-BABA tests allows such analyses to be done on datasets that span deeper evolutionary time. While gene tree-based ABBA-BABA tests have been utilized in the literature (e.g. the chi-squared test in (Suvorov et al. 2021)), to date there are no software packages for performing these analyses.

Here we present an R package, *quaint*, that detects introgression from phylogenomic datasets by leveraging gene tree topologies rather than nucleotide site patterns. By adapting the ABBA-BABA framework to use alternative gene tree topologies, *quaint* overcomes key disadvantages of site-based methods. This package automates the testing of all possible taxon quartets across large phylogenies, while providing summary measures of the results of individual hypothesis tests for easier interpretation. This software package will be a valuable tool for phylogenomic researchers, increasing the accessibility and reproducibility of gene tree-based ABBA-BABA tests for introgression.

## METHODS AND RESULTS

Using a framework similar to the ABBA-BABA test (Green et al. 2010), *quaint* tests whether or not introgression between two taxa has occurred. As with the ABBA-BABA test, *quaint* uses quartets of four taxa where the relationships are (O,(P3,(P2,P1))). Given this set of relationships, there are three possible topologies: one that is concordant with the species tree, and two that are discordant with the species tree (Figure 1). Under a null coalescent model (i.e. with only ILS contributing to gene tree discordance), the two discordant topologies will occur in equal proportions, while introgression between two of the taxa in the quartet will cause one of the discordant topologies to exceed the frequency of the other (Figure 1). *quaint* uses a chi-square goodness of fit test to determine if one of the two discordant topologies is statistically greater than the other. Additionally, a statistic similar to Patterson's D-statistic, $D_q$, is calculated to quantify how skewed the topologies are.

$$D_q(P1, P2, P3, O) = \frac{ABBA - BABA}{ABBA + BABA}$$

Typical datasets will have many more than four taxa, meaning that there is often more than one quartet of taxa that can be used to calculate $D_q$ for a given pair of taxa. In these cases, *quaint* provides a summary statistic which takes the mean $D_q$ of all quartet tests for a given taxon pair, setting any $D_q$ values calculated from a non-significant quartet test to zero. Additionally, any quartet tests with negative $D_q$ values are set to zero. This summary statistic provides a robust representation of the support for introgression between any two taxa in the provided dataset.

## CONCLUSIONS

*quaint* provides an accessible and reproducible framework for detecting introgression in phylogenomic datasets. This package circumvents the limitations of site-based methods for detecting introgression by implementing a gene tree-based ABBA-BABA test, enabling the detection of introgression in deeper evolutionary time. *quaint* will serve as a valuable tool for evolutionary biologists grappling with the complexity of reticulate evolution across the tree of life.

## ACKNOWLEDGEMENTS

We would like to thank Nick Chang for testing the functionality of *quaint* and discovering bugs before release. We also would like to thank Philip Bentz and Summer Blanco for providing data that assisted in testing the functionality of the software. Funding provided by the Rosemary Grant Advanced Award from the Society for the Study of Evolution.

## AUTHOR CONTRIBUTIONS

E.A.B. and J.L.-M. conceptualized the project. E.A.B. wrote the code and wrote the initial manuscript draft. J.L.-M. revised and edited the manuscript.

## DATA AVAILABILITY STATEMENT

Baldwin and Leebens-Mack – An R Package for detecting introgression.

The code for the R package is available on GitHub (https://github.com/ethan-baldwin/quaint). The code used at the time of writing, is commit e786a4f.

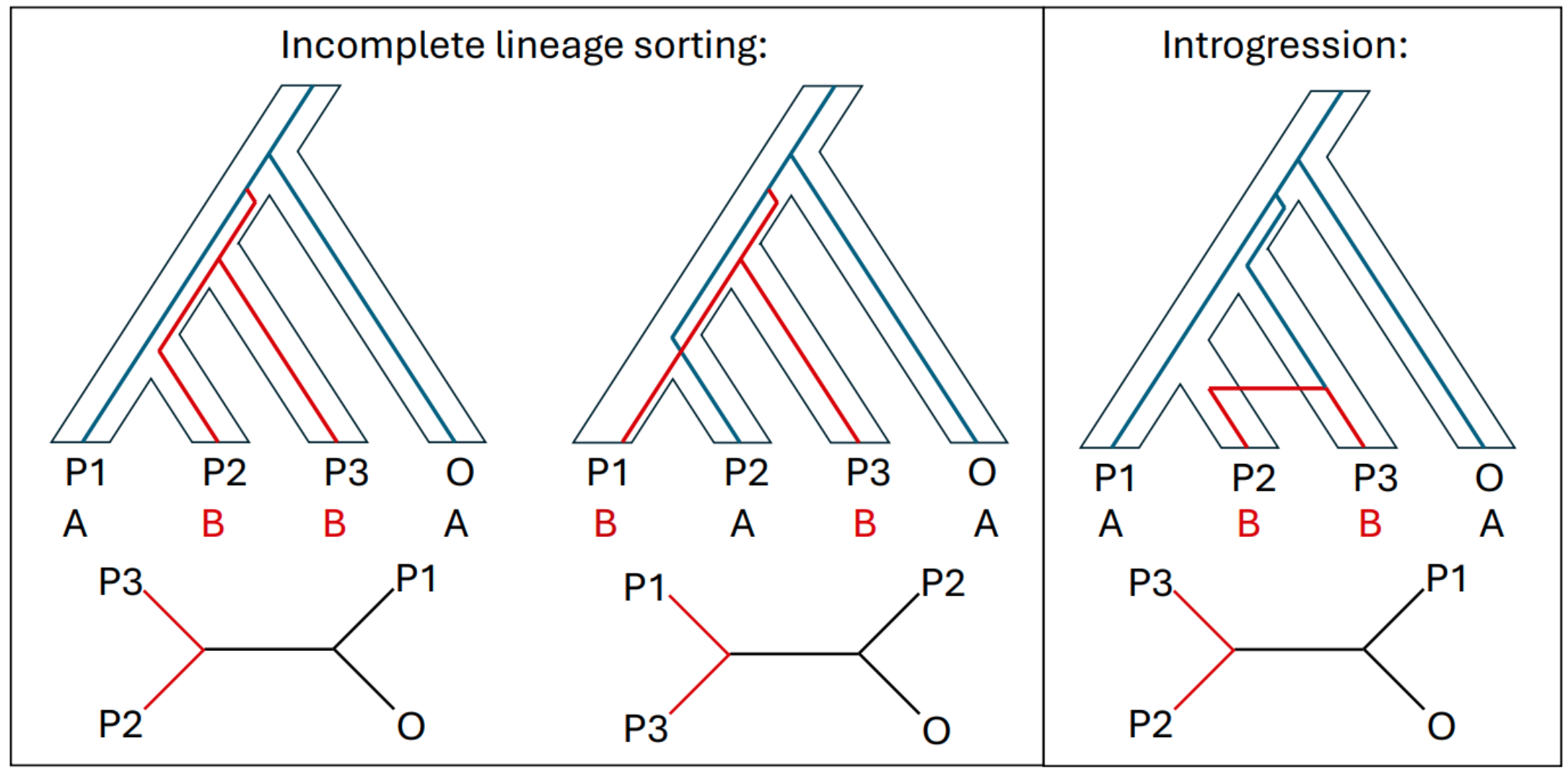

Figure 1. Theoretical framework behind the ABBA-BABA test. The top three trees demonstrate different scenarios wherein ABBA and BABA site patterns or gene tree topologies are generated. Incomplete lineage sorting will produce ABBA and BABA topologies in equal proportions, while introgression between P2 and P3 produces only ABBA topologies. Below each scenario is the resulting unrooted gene tree topology.